**Self-assembled nano-columns in Bi$_2$Se$_3$ grown by molecular beam epitaxy**

*Theresa P. Ginley and Stephanie Law\**

Department of Materials Science and Engineering, University of Delaware, 127 The Green Room 201, Newark DE 19716, United States of America

*Author to whom correspondence should be addressed: slaw@udel.edu

**Abstract**

Layered van der Waals (vdW) materials grown by physical vapor deposition techniques are generally assumed to have a weak interaction with the substrate during growth. This leads to films with relatively small domains that are usually triangular and a terraced morphology. In this paper, we demonstrate that Bi$_2$Se$_3$, a prototypical vdW material, will form a nano-column morphology when grown on GaAs(001) substrates. This morphology is explained by a relatively strong film/substrate interaction, long adatom diffusion lengths, and a high reactive selenium flux. This discovery paves the way toward growth of self-assembled vdW structures even in the absence of strain.

**I. Introduction**

Layered van der Waals (vdW) materials such as the topological insulator (TI) Bi$_2$Se$_3$ are a highly anisotropic family of materials that have strong bonds in the *a-b* plane and weak vdW bonds along the *c*-direction. The materials in this family have a variety of potential applications including next-generation optics,[1] optoelectronics,[2] spintronics,[3,4] and valleytronics.[5] vdW materials can be synthesized as bulk crystals and exfoliated into thin sheets or grown directly on a variety of substrates using chemical vapor deposition, molecular beam epitaxy (MBE), or related techniques. To date, the majority of efforts have focused on growing large-area films of these materials in the (0001) orientation.[6,7] Films in this orientation have the vdW gaps oriented parallel to the surface of the substrate. For most applications, smooth and flat films with coalesced domains would be ideal. Unfortunately, vdW films generally show relatively small domains with pyramidal or "wedding cake" morphologies and a large amount of twinning. MBE growth of related material systems, such as transition metal dichalcogenides (a popular two-dimensional material), show similar morphologies.[8–10] The common morphology of vdW materials can be attributed to the weak film-substrate interaction, which causes domains to nucleate with different rotational orientations and for subsequent layers to start growing before the first layer has finished. The weak interaction also causes vdW films to predominantly grow in the (0001) orientation. Alternative surface orientations for vdW materials have been obtained through substrate pre-patterning or surface oxidation, which changes the film-substrate interaction.[11,12]

In addition to controlling film morphology and surface orientation, film-substrate interactions have been used in other systems to grow self-assembled nanostructures. In semiconductor materials, quantum dots have been created using strain-driven self-assembly techniques, and nanowires have been grown using the vapor-liquid-solid mechanism in which gold nanoparticles are used to seed nanowire growth.[13,14] Due to the weak interaction between the film and substrate, strain-driven self-assembly of nanostructures is unlikely to succeed in TI or other vdW materials. To date, TI nanowires have only been synthesized through the vapor-liquid-solid method or by templated growth.[15–18]

In this paper, we report the growth of columnar nanostructures of $Bi_2Se_3$ on GaAs(001) substrates. We show that the presence and density of the columnar features can be controlled by varying the substrate temperature and the growth:anneal (G:A) time ratio used during growth. These results demonstrate the importance of the film-substrate interaction even when growing vdW materials and open the door to new TI morphologies. Much like quantum dot growth unlocked new device physics in semiconductors, control over the growth of TI nano-columns would enable the exploration of new phenomena in topologically non-trivial materials in the quantum regime. This result also demonstrates the ability to grow self-assembled nanostructures in vdW materials without using strain-driven assembly techniques.

## II. Results

All samples were grown by MBE in a dedicated Veeco GenXplor reactor on either *c*-plane Al$_2$O$_3$ or undoped GaAs(001) substrates. A valved selenium cracker source was used for all deoxidizations and growth.[19] All GaAs(001) substrates were first outgassed in the load lock at 200°C for 12 hours and then transferred into the growth chamber and heated to 770°C in a selenium overpressure to thermally desorb the GaAs oxide layer. The selenium valve was opened after the substrate reached 300°C. The Al$_2$O$_3$ substrates were outgassed at 200°C for 12 hours in the load lock then heated to 650°C in the growth chamber to desorb any remaining surface contaminants. A temperature of 900°C was used for the selenium cracking zone while the flux was controlled using the valve. A multistage growth process was used for all films as described in Yong et al.[20] First, a 5 nm layer of Bi$_2$Se$_3$ was deposited at 325°C followed by a 5nm layer of In$_2$Se$_3$. This 10nm layer is the seed layer and its composition is the same for all growths. The seed layer was then heated to 425°C and annealed for 40 min. During the annealing process, the bismuth and indium interdiffuse, forming a (Bi$_{0.5}$In$_{0.5}$)$_2$Se$_3$ alloy.[21,22] The total growth rate was ~0.8 nm/minute throughout all stages of growth. The seed layer was grown using the following cycle: one-minute deposition followed by one-minute anneal. Bi$_2$Se$_3$ was then deposited on the seed layer for 40 minutes of active growth time. During this stage the growth:anneal (G:A) ratio and the substrate temperatures were varied between samples. All substrate temperatures were measured using a non-contact thermocouple. All fluxes reported are beam equivalent pressures measured via a beam flux monitor. To prevent selenium outgassing, the selenium overpressure was sustained after growth until the substrate temperature dropped below 200°C. Atomic force microscopy (AFM) was used to determine the surface morphology of the samples. When placed in the AFM, the samples were aligned by eye using the major flat. The inexact nature of this alignment may account for any perceived deviation from the film alignment to the substrate discussed in this paper.

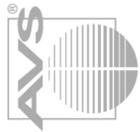

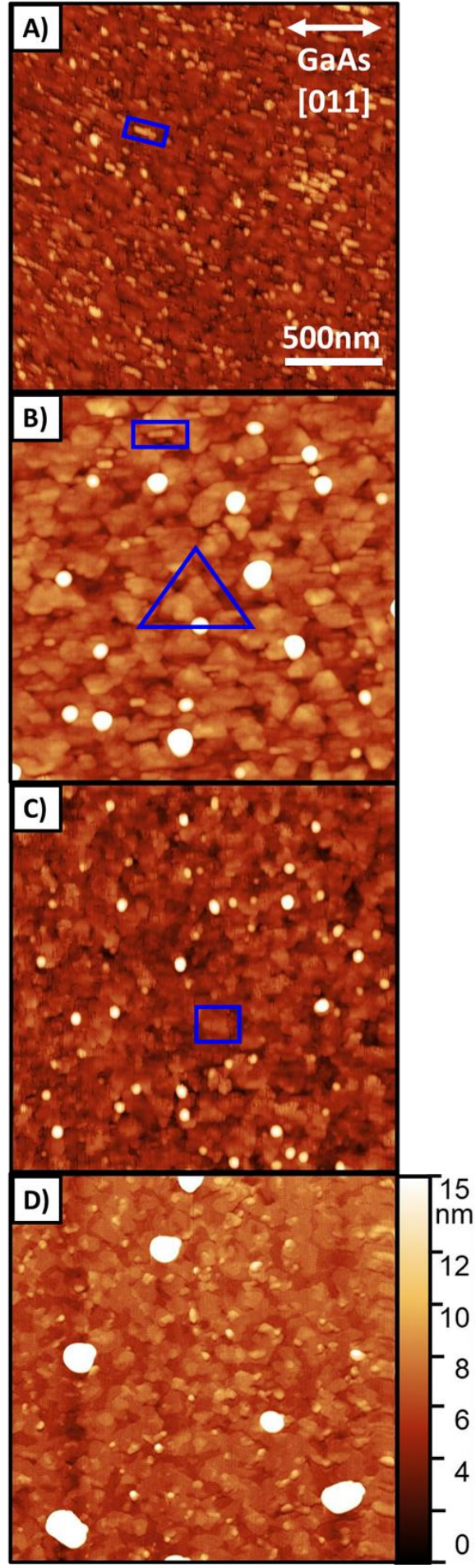

**FIG 1.** AFM images of A) 5nm of $Bi_2Se_3$ on GaAs(001), B) the full seed layer (5nm of $Bi_2Se_3$ followed by 5nm of $In_2Se_3$) on GaAs, C) the full seed layer after a 40 minute anneal on GaAs (001) and D) the annealed seed layer on AlOx. The scale bar is the same for all samples. The substrate orientation for samples A-C is nominally the same, however the orientation in the AFM was done by eye so may be inexact. Blue outlines for key features have been added as a guide to the eye. The tall white dots are attributed to dust on the films, not film morphology.

We will first discuss how the seed layer nucleates on the GaAs(001) substrate as compared to $AlO_x$ substrates. Understanding the nucleation of this layer sets the stage for discussing the unique morphologies we achieve. A series of three films (Samples A-C) were grown representing each stage of the seed layer. Sample A was terminated immediately after the initial 5nm of $Bi_2Se_3$, Sample B was terminated after the 5nm of $Bi_2Se_3$ and 5nm of $In_2Se_3$, and Sample C was terminated after the anneal. As shown in Fig. 1(A), after the growth of the initial 5 nm of $Bi_2Se_3$, the film has formed small rectangular domains extending along the [011] direction of the substrate. The disparity between the marked substrate orientation and the feature alignment in Fig. 1(A) is most likely due to inexact alignment in the AFM. A blue box in Fig. 1(A) highlights one of these features. In Sample B shown in Fig. 1(B), the rectangular domains shown in the AFM image have spread into triangular domains more typical of this material system, as highlighted by the blue triangle. It is important to note that the newly formed triangular domains inherit the orientation to the substrate established in the $Bi_2Se_3$ layer. For Sample C, the domains in the AFM image in Fig. 1(C) have become more cohesive while maintaining their orientation to the substrate as highlighted by the blue box. While the orientation to the substrate has weakened as the domains shifted during the anneal stage, it is still evident. Fig. 1(D) shows a completed seed layer grown on an AlOx substrate; no substrate orientation is apparent in the image and the domains appear more rounded than their counterparts grown on GaAs(001). The tall white dots on the images are attributed to dust on the samples, not features on the films.

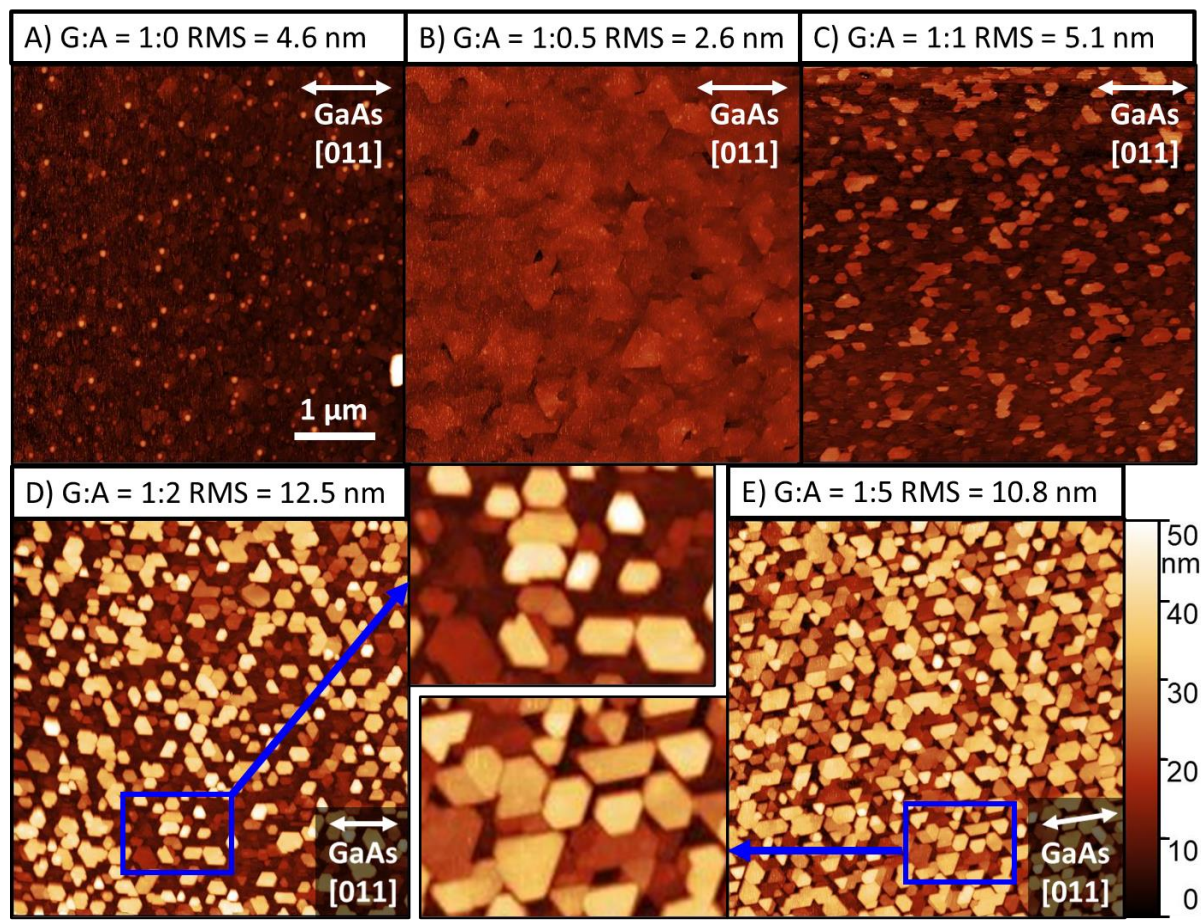

**FIG 2.** AFM scans of Bi$_2$Se$_3$ films grown with varying G:A time ratios on GaAs(001) substrates. All films were grown on a 10nm BiInSe$_3$ seed layer for a cumulative growth time of 40 minutes. The scale is the same for all samples. The substrate orientation is nominally the same for all samples, however the orientation in the AFM was done by eye so may be inexact. Magnifications of D and E have been included to better show the orientation of the features to the substrate.

Next, we studied the effect of the grow:anneal (G:A) time ratio on the morphology of films grown on the complete 10nm BiInSe$_3$ seed layer on GaAs(001) using a substrate temperature of 425°C. A series of five films were grown with G:A ratios varying from 1:0 (continuous growth) to 1:5 (growth for 1 minute then anneal for 5 minutes). In all cases, the growth time was kept at 1 minute and the anneal time changed. The film grown with no annealing (Fig. 2(A)) shows very small domains. Annealing for 30 seconds every cycle (Fig. 2(B)) resulted in a smooth film with the triangular domains traditionally observed in van der Waals epitaxy. With anneal times of 1 minute and above (Fig. 2(C-E)) we see the emergence of a three-dimensional morphology. These AFM images show hexagonal columnar nanostructures. As the anneal time is increased, the in-plane dimensions of these structures remain relatively constant while their height and density increase, as can been seen by both the color scale and the increased RMS roughness in the 1:2 and 1:5 samples. In all the films, apart from the 1:0 sample, we see alignment along the GaAs [011] direction. Magnificantions of the AFM scans for the samples with 1:2 and 1:5 G:A ratios are shown in Fig. 2 to highlight this substrate alignment.

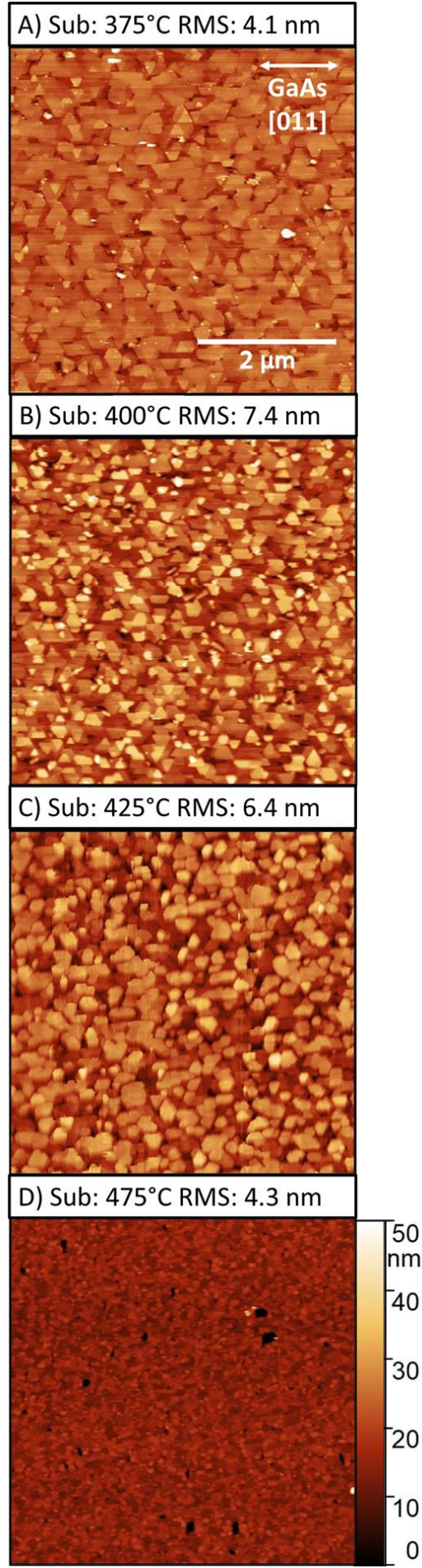

**FIG 3.** AFM scans of $Bi_2Se_3$ films grown on GaAs(001) substrates with varying substrate temperatures. All films were grown with a G:A ratio of 1:1 on a 10nm $BiInSe_3$ seed layer for a cumulative growth time of 40 minutes. All scans are shown with the same scales and substrate orientation.

Fig. 3 shows the effect of substrate temperature on film morphology for samples grown on GaAs(001) using a G:A ratio of 1:1. At a substrate temperature of 375°C (Fig. 3(A)) we see a relatively smooth film with mostly triangular domains. Increasing the substrate temperature to 400°C and 425°C (Figs. 3(B) and 3(C)) results in a transition to the columnar morphology. At a substrate temperature of 475°C, surface adatoms reevaporate readily and we see only a small amount of film growth. All four temperatures exhibit film alignment along the GaAs[011] direction.

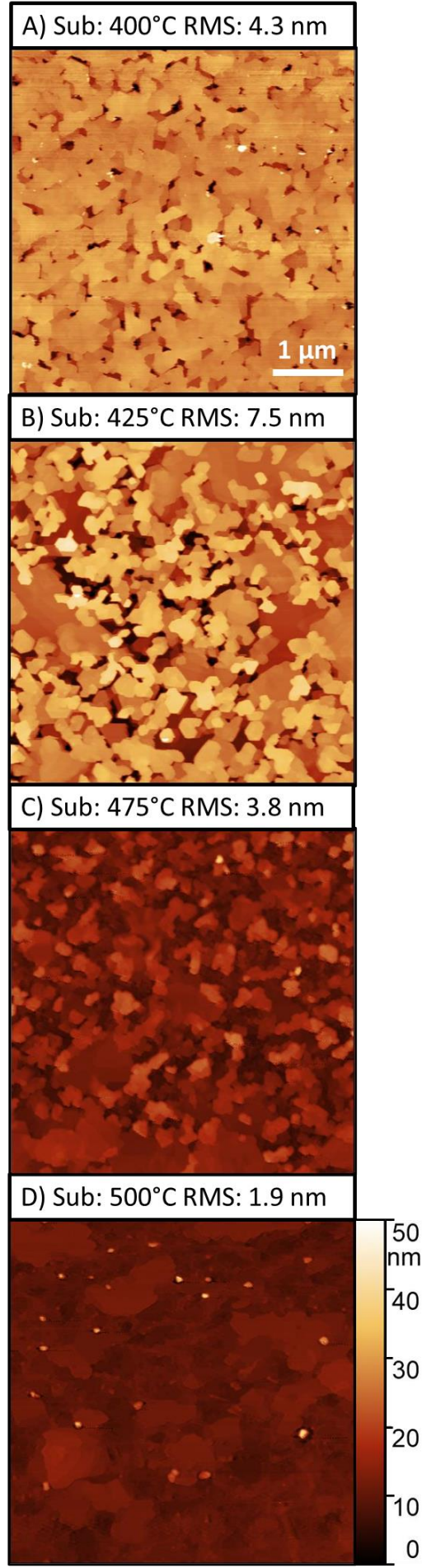

**FIG 4.** AFM scans of Bi$_2$Se$_3$ films grown with varying substrate temperatures on Al$_2$O$_3$ substrates. All films were grown on a 10nm BiInSe$_3$ seed layer with a 1:1 G:A ratio for a cumulative growth time of 40 minutes. All scans are shown with the same scales.

Finally, Fig. 4 shows the effect of substrate temperature on films grown on Al$_2$O$_3$ substrates with a G:A ratio of 1:1. It is notable that all temperatures used in this study are above the thermal degradation point for films grown on sapphire without the seed layer. None of the AFM images in Fig. 4 show the triangular domains we see in growth on Al$_2$O$_3$ at lower temperatures. However, we also do not see the distinct and separate columns seen in high temperature growth on GaAs(001). Instead, the samples display fractal like-domains[23] (most obvious in Fig. 4(B)). Increasing the temperature further reduces adatom incorporation and thus limits film growth. Also unlike films grown on GaAs(001), none of the films demonstrate significant substrate alignment.

### III. Discussion

We will now discuss how the interaction between the film and the substrate is driving the formation of these unique morphologies. We first discuss the morphological evolution of the seed layer. When growing the seed layer on GaAs(001), we see a strong substrate alignment and anisotropic elongation along the GaAs[011] direction emerging during the initial Bi$_2$Se$_3$ deposition step, Fig. 1(A). During the subsequent In$_2$Se$_3$ growth and annealing process the domains become less elongated yet remain aligned to the substrate. This behavior is markedly different from the random domain orientation we observe in seed layer growth on sapphire substrates.[20] One major difference between a sapphire substrate surface and that of GaAs(001) is the anisotropy in surface energy due to the (2x4) GaAs surface reconstruction after the deoxidization process. We, therefore, attribute the initial alignment and domain elongation to the preferential fast diffusion of bismuth along the GaAs[011] direction.[24]

In order to understand the columnar growth morphology, we must first discuss how TI films normally grow. The hexagonal unit cell of these materials has two distinct edges in the *a-b* plane; $[\bar{1}2\bar{1}]$, which has two dangling bonds per edge atom, and $[\bar{2}11]$, which only has one. This results in faster growth along the $[\bar{1}2\bar{1}]$ edges due to preferential atom incorporation.[25] The difference in growth rates is so large that it results in the characteristic triangular domain shape typically seen in $Bi_2Se_3$ and $In_2Se_3$ growth instead of a hexagonal domain.[19,20,26] Additionally, modeling has shown that as the substrate-film interaction weakens, the energy barrier that prevents adatoms from hopping up decreases.[23] In vdW materials the inherently weak substrate-film interactions allows for significant adatom diffusion up step edges, resulting in a terraced film morphology. These two mechanisms combine to produce the pyramidal triangular domains characteristic of vdW films.

These growth mechanisms do not adequately explain the nanostructured growth reported in this paper. Instead, we can make a comparison to the catalyst-free growth of GaN nanowires.[27–32] At high substrate temperatures and low gallium fluxes, GaN has been demonstrated to form nanowires during MBE growth. Nanowire nucleation density can be increased by using atomic nitrogen instead of less reactive nitrogen dimers, or by increasing the V:III ratio.[29] In the initial stage of growth, gallium adatoms randomly diffuse along the surface until they aggregate and nucleate domains. Nucleation density is therefore controlled by the gallium diffusion length.[33] Gallium adatom incorporation is much more energetically favorable on the tip of the nanowire than on the side walls.[33] An adatom that impinges on the side wall must diffuse to the tip to incorporate, resulting in the nanowire growth perpendicular to the substrate surface.[27] However if the distance to the tip is greater than the gallium diffusion length, the adatom will instead re-evaporate.[30] This means that the formation of nanowires requires a long gallium diffusion length.[28] To achieve the columnar $Bi_2Se_3$ morphology reported in this paper we use a very high flux of reactive (cracked) selenium molecules, and we need conditions that increase bismuth adatom mobility. Since these conditions mimic those of GaN nanowire growth and the morphology of the $Bi_2Se_3$ samples is similar to GaN nanowires, we propose that the same mechanisms can be used to understand both material systems.

One point that still requires explanation is why bismuth adatoms no longer incorporate readily at domain side walls, like they do in the more common triangular $Bi_2Se_3$ growth morphologies. For nanocolumns to grow, adatoms must preferentially incorporate at the top of the nanocolumn, rather than on its side. We observe that most of the nanocolumns are hexagonal or rhombohedral, not exclusively triangular as observed for the domains in traditional $Bi_2Se_3$ films. This suggests that the previously discussed difference in growth rates between the $[\bar{1}2\bar{1}]$ and $[\bar{2}11]$ edges is no longer regulating the domain shape. We propose that the large selenium overpressure we use during growth is causing a reconstruction of the dangling selenium bonds along the nanowire side walls (much like the change in III-V surface reconstructions as a function of group V overpressure). The reconstruction equalizes the number of dangling bonds on each edge, resulting in more isotropic growth in the *a-b* plane. Additionally, reduced dangling bonds on the side walls would reduce bismuth adatom incorporation, explaining the GaN nanowire like growth dynamic. [34–37] When we use lower selenium overpressures but otherwise similar growth conditions, we do not observe the nanocolumn morphology[38].

The dependence of columnar behavior on bismuth diffusion is well supported by the AFM images in Figs. 2 and 3. In Fig. 2, we show that increasing the anneal time between growth steps increases the height and density of the columnar features. During the anneal steps, the bismuth adatoms will have more time to diffuse. Bismuth that would otherwise be buried by further deposition can instead diffuse far enough to incorporate into existing columns or form new ones. If the anneal time is too short (Fig. 2(A) and 2(B)), the bismuth adatoms will not have enough time to diffuse and form columns and a more traditional morphology will appear. This trend is also seen when we look at the effect of substrate temperature during growth on GaAs(001), shown in Fig. 3. At low temperatures, the adatom mobility is reduced and we see a traditional morphology. Increasing the substrate temperature, and thus the adatom mobility, allows the columnar growth mode to emerge.

The differences in morphology between films grown on GaAs(001) and films grown on $Al_2O_3$ highlight the importance of the film/substrate interaction. During the growth of the $Bi_2Se_3$ portion of the seed layer on GaAs(001) substrates, we see a strong alignment along the [011] axis of the substrate which we attribute to the fast diffusion of bismuth adatoms along this direction. This initial orientation to the substrate is propagated through all growth stages, culminating in an alignment of the nanocolumns along the same axis. Unlike GaAs(001), $Al_2O_3$ provides an isotropic growth surface for the $Bi_2Se_3$ so no preferred orientation to the substrate is generated. While we still observe a high degree of three dimensionality in the films grown on $Al_2O_3$ (Fig. 4(B)), the edges of the features are much more fractal than the hexagonal features grown on GaAs(001) substrates. We attribute this to the lack of preferred orientation to a weaker interaction with the substrate.

## 4. Conclusion

In this paper, the growth of $Bi_2Se_3$ on GaAs(001) using a $(Bi_{0.5}In_{0.5})_2Se_3$ seed layer was compared to growth on $Al_2O_3$ substrates with seed layers. It was found that the presence of the seed layer allowed for growth at much higher temperature than previously achieved on $Al_2O_3$ substrates. At these high temperatures. a three-dimensional morphology emerged on both substrate types. However, the fast diffusion of bismuth along the GaAs [110] direction during the seed layer growth resulted in a preferred orientation to the substrate that was maintained during the rest of the growth. This resulted in the growth of confined oriented nanocolumn structures. While this morphology may be unexpected for a van der Waals system, can be understood by analogy with the growth of catalyst-free GaN nanowires. This result points to the importance of the film/substrate interaction even in normally weakly-interacting vdW materials and demonstrates the ability to synthesize self-assembled nanostructures in vdW materials without using strain-driven assembly. The discovery of MBE-grown vdW material morphologies with self-assembled nanostructures may allow researchers to better study quantum phenomenon in topological insulators. Much like controlled growth of semiconductor quantum dots, further understanding and control of this growth morphology could unlock exiting topological insulator research and technology.

**Acknowledgements**

T. P. G. and S. L. acknowledge funding from the U.S. Department of Energy (DOE), Office of Science, Office of Basic Energy Sciences under Award No. DE-SC0017801. T. P. G. and S. L. acknowledge the use of the Materials Growth Facility (MGF) at the University of Delaware, which is partially supported by the National Science Foundation Major Research Instrumentation under Grant No. 1828141 and is a member of the Delaware Institute for Materials Research (DIMR).

**Data availability**

Data available in article.

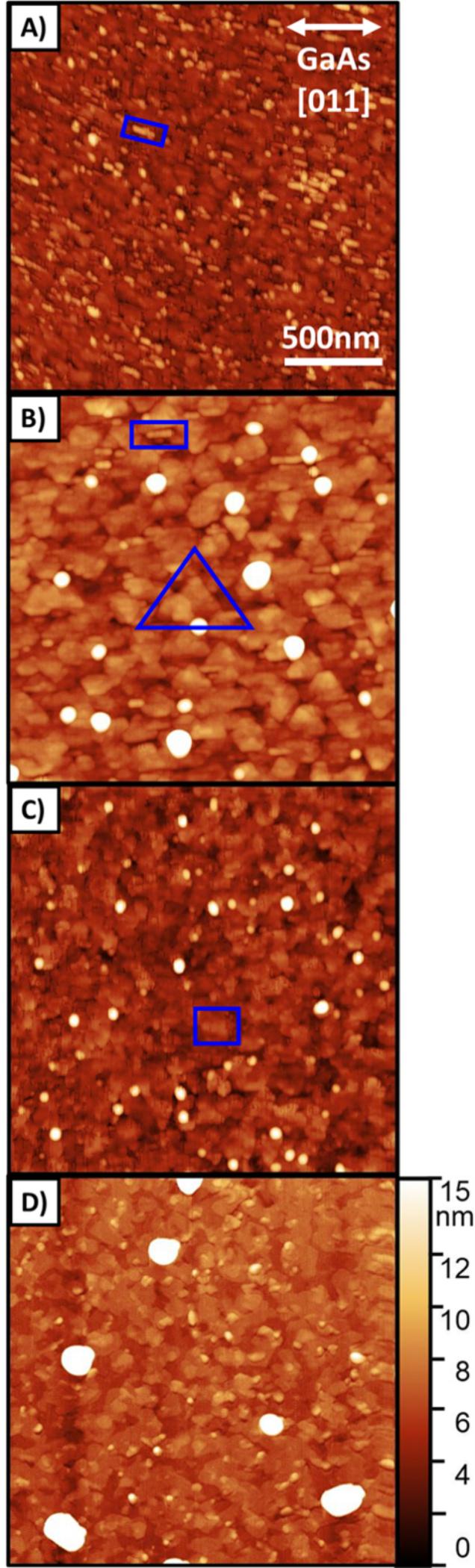

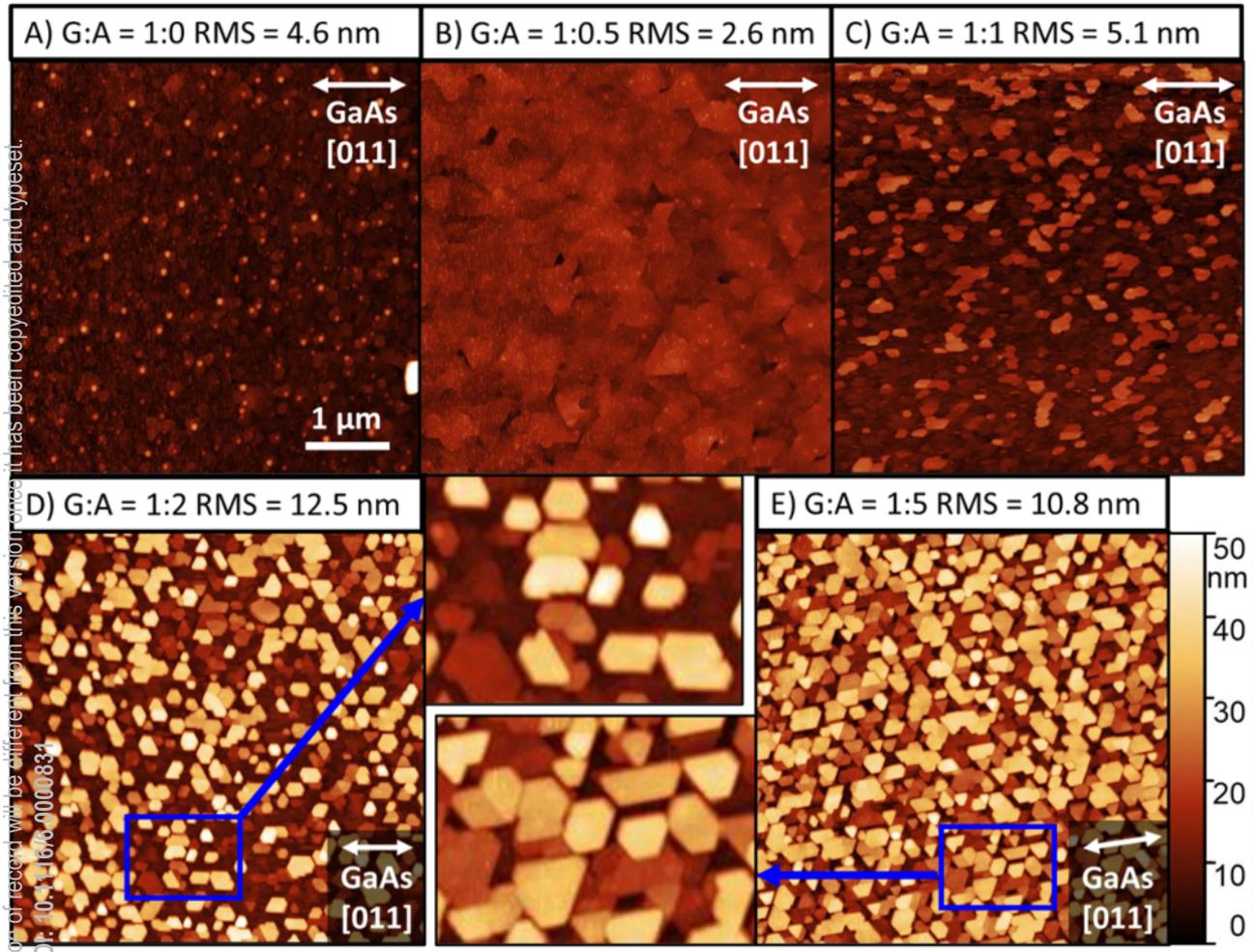

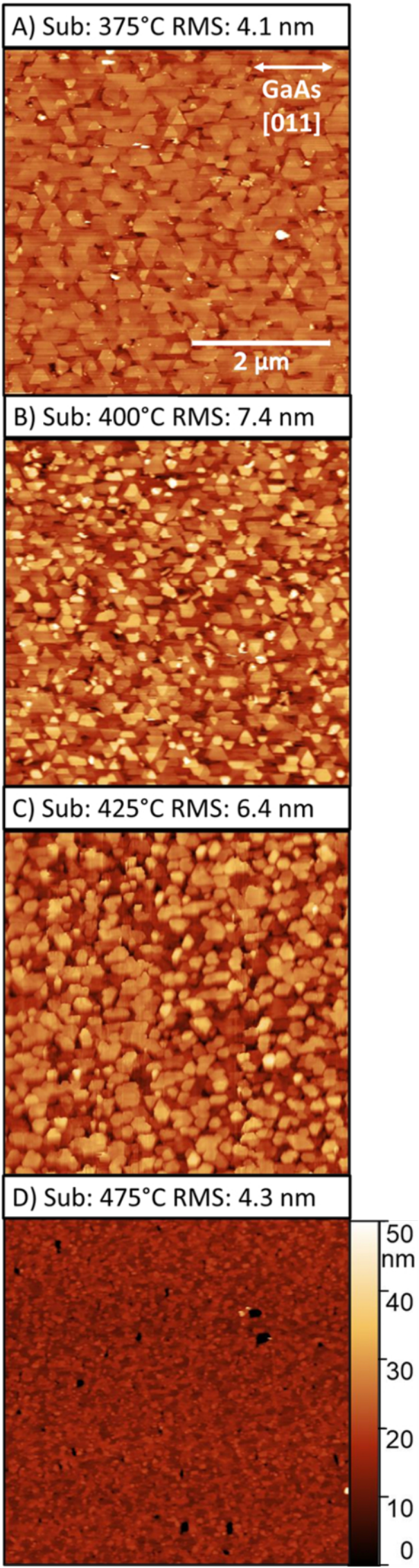

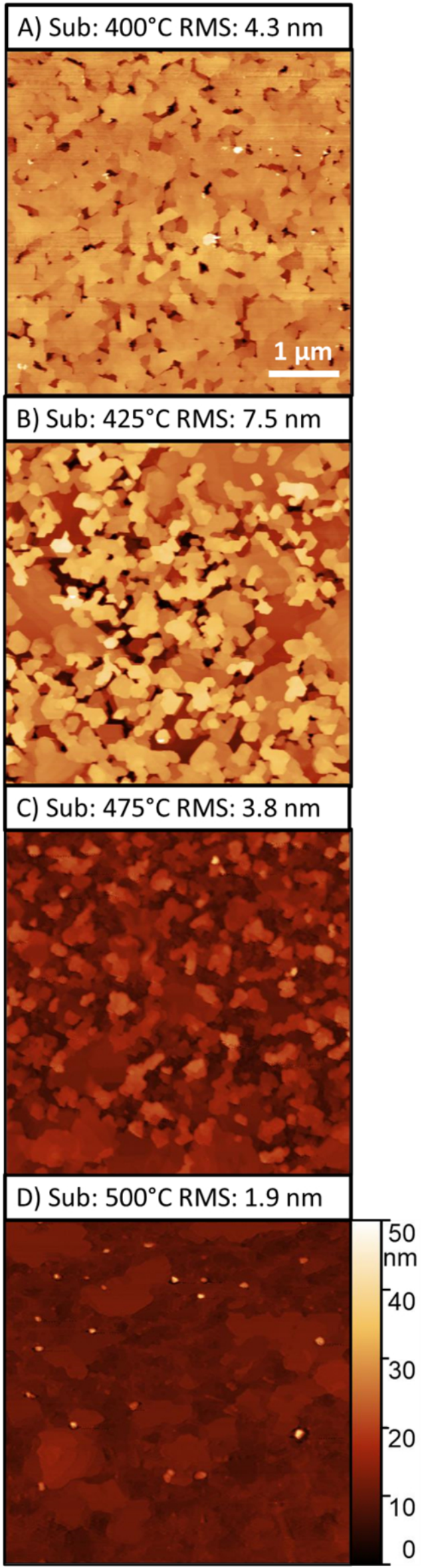